\documentclass[superscriptaddress,amsmath,amssymb,aps,reprint,prb,twocolumn]{revtex4-2}
\usepackage[dvipdfmx]{graphicx}
\usepackage{amsmath,bm,amssymb}
\usepackage{hyperref}
\usepackage{float}
\usepackage{subfigure}
\usepackage[english]{babel}
\usepackage{comment}
\usepackage{physics}

\newcommand{\li}{\begin{aligned}}
\newcommand{\eli}{\end{aligned}}
\newcommand{\be}{\begin{equation}}
\newcommand{\ee}{\end{equation}}

\usepackage{color}

\begin{document}

\title{Microscopic derivation of the microstretch theory for carbon nanotubes}

\author{Naoki Nishimura}
\affiliation{Institute for Solid State Physics, University of Tokyo, Kashiwa, 277-8581, Japan}

\author{Mamoru Matsuo}
\affiliation{Kavli Institute for Theoretical Sciences, University of Chinese Academy of Sciences, Beijing, 100190, China.}
\affiliation{CAS Center for Excellence in Topological Quantum Computation, University of Chinese Academy of Sciences, Beijing 100190, China}
\affiliation{Advanced Science Research Center, Japan Atomic Energy Agency, Tokai, 319-1195, Japan}
\affiliation{RIKEN Center for Emergent Matter Science (CEMS), Wako, Saitama 351-0198, Japan}

\author{Takeo Kato}
\email{kato@issp.u-tokyo.ac.jp}
\affiliation{Institute for Solid State Physics, University of Tokyo, Kashiwa, 277-8581, Japan}

\date{\today}

\begin{abstract}
Twisted carbon nanotubes support phonons involving not only torsion, naturally associated with microrotation, but also radial breathing, which requires a scalar stretch degree of freedom. We derive an effective microstretch theory for these modes starting from nonlinear elasticity on a cylindrical surface. By linearizing the equation of motion around a uniformly twisted equilibrium configuration, we obtain the dynamical matrix for the twisting, longitudinal, and radial-breathing modes. This matrix coincides with that of a one-dimensional microstretch theory, and the corresponding elastic constants are expressed in terms of the Lam\'e constants, the nanotube radius, and the twist rate. The twist generates chiral couplings in the effective theory, which hybridize the three modes and open an anticrossing in the phonon dispersion. These results provide a microscopic basis for the microstretch description of phonons in twisted carbon nanotubes and clarify how structural chirality enters the effective couplings.
\end{abstract}
    
\maketitle

\section{Introduction}

The microscopic description of phonons in chiral materials has recently emerged as a central topic in solid-state physics~\cite{ZhangNiu2014,Nakane2018,Hamada2018,Park2020,Hamada2020,Streib2021}. Modern formulations of phonon angular momentum have drawn increasing attention to how structural handedness affects phonon polarization, mode mixing, and related dynamical properties~\cite{Zhang2015,Zhu2018,Ishito2023,Chen2021,Tsunetsugu2023,Ohe2024,Funato2024,YangRen2024,Nishimura2025,Juraschek2025}. Carbon nanotubes provide a particularly useful platform for this problem~\cite{Saito1998,Suzuura2002,Jorio2005,GuptaKumar2019,Akimoto2026}: their cylindrical geometry supports well-defined twisting, longitudinal, and radial-breathing modes, while an externally imposed twist offers a controllable source of handedness. Twisted nanotubes therefore provide a transparent and tunable setting in which to examine how structural chirality is encoded in low-energy phonon dynamics.

An effective continuum description of these modes must account not only for torsional motion, which is naturally associated with local rotation, but also for the radial-breathing mode, which involves a change in tube radius. This means that the relevant low-energy dynamics requires not only translational and rotational variables but also an additional scalar stretch-like degree of freedom. Such a situation is naturally described within the framework of generalized continuum elasticity. In particular, micropolar elasticity~\cite{Nowacki1985,Eringen2012,Kishine2020,Matsuo2026} extends conventional elasticity by endowing each material point with an independent rotational degree of freedom in addition to its translational displacement, whereas microstretch theory further introduces a scalar dilatational degree of freedom associated with local isotropic deformation. For carbon nanotubes, the twisting mode is naturally identified with microrotation, while the radial-breathing mode requires an additional scalar stretch variable. The low-energy phonon dynamics of twisted carbon nanotubes is therefore more naturally captured by a one-dimensional microstretch framework.

An effective one-dimensional microstretch description of carbon-nanotube phonons has indeed been shown to capture characteristic features of the dispersion relation obtained in molecular-dynamics calculations~\cite{GuptaKumar2019}. However, the elastic coefficients were introduced phenomenologically there, and their microscopic origin has remained unclear. In particular, it has not been established how a twisted equilibrium configuration generates the chiral coupling terms that mix torsional, longitudinal, and radial-breathing modes. Clarifying this point is essential for connecting structural chirality in carbon nanotubes to the effective continuum description of their phonon dynamics.

\begin{figure}[tb]
    \centering
    \includegraphics[width=7cm]{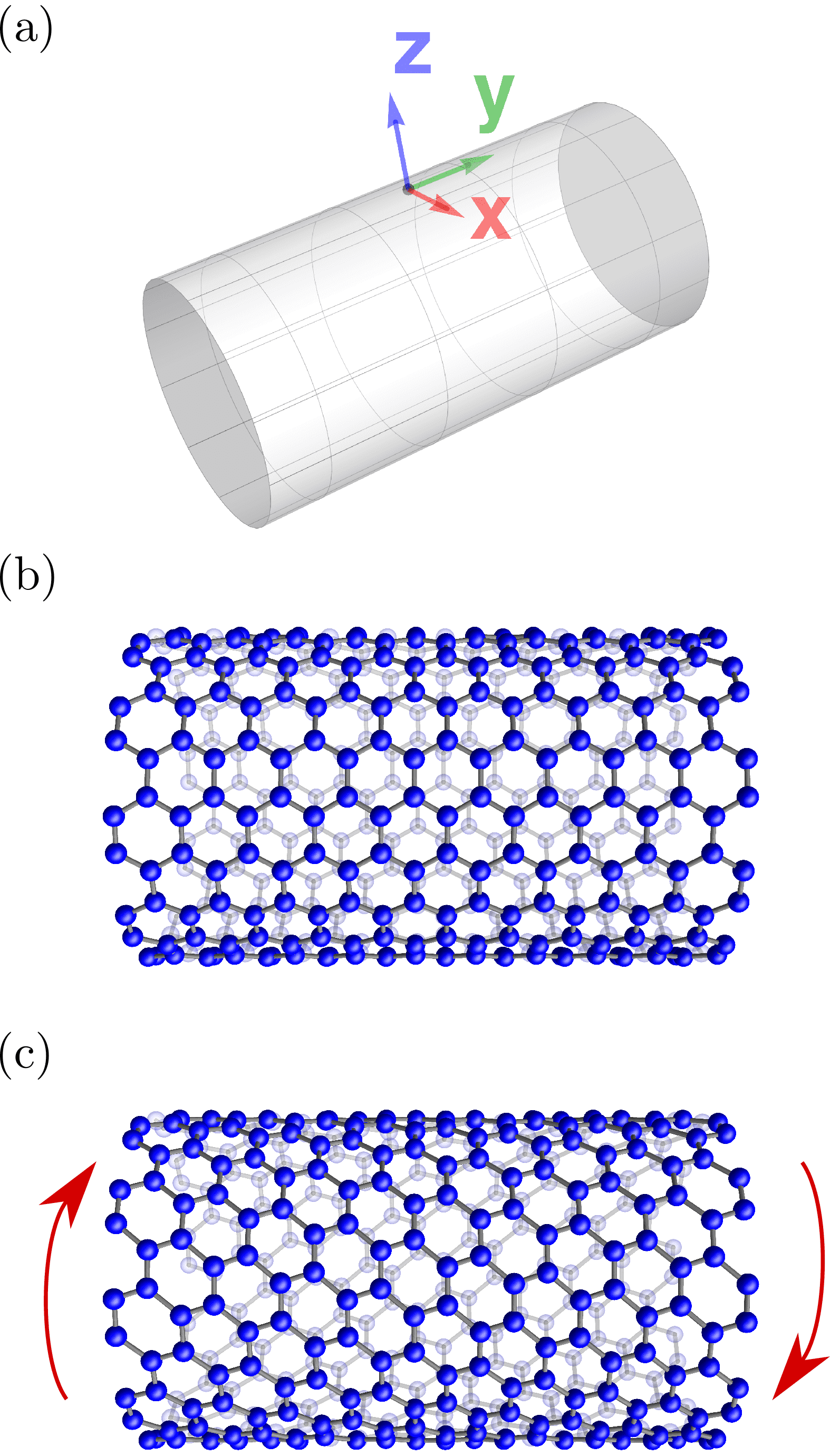}
    \caption{(a) The coordinate system of nanotube.
    (b) Stress-free configuration.
    (c) Twisted configuration.}
    \label{coord_initial_twist}
\end{figure}

In this work, we provide a microscopic derivation of the effective one-dimensional microstretch description for phonons in twisted carbon nanotubes (see Fig.~\ref{coord_initial_twist}). Starting from nonlinear elasticity on a cylindrical surface, we linearize the equation of motion around a uniformly twisted static configuration and derive the dynamical matrix for the twisting, longitudinal, and radial-breathing modes. In particular, we show that the chiral coupling terms of the effective theory arise from the twisted equilibrium configuration through the linearization of the nonlinear elastic force. Within an isotropic elastic approximation, we further obtain analytic expressions for the effective coefficients in terms of the Lam\'e constants, the nanotube radius, and the twist rate. Our derivation thereby clarifies the microscopic origin of the chiral couplings and shows how structural chirality is encoded in the effective phonon dynamics.

The remainder of this paper is organized as follows. In Sec.~\ref{sec:microstretch}, we summarize the effective one-dimensional microstretch theory relevant to carbon nanotubes. In Sec.~\ref{sec:model}, we formulate a microscopic nonlinear elastic theory for a twisted nanotube and derive the linearized equation of motion around the twisted equilibrium state. We also obtain the dynamical matrix, identify the corresponding elastic constants in the effective theory, and discuss the resulting dispersion relation. Finally, we summarize our findings in Sec.~\ref{sec:summary}.

\section{Effective one-dimensional microstretch theory}
\label{sec:microstretch}

We first summarize an effective one-dimensional microstretch theory for a rod, which was shown to be consistent with the phonon dispersion of carbon nanotubes obtained from molecular-dynamics calculations~\cite{GuptaKumar2019}.
In standard elasticity theory, one considers the displacement field $\vb*{u}({\bm r})$ of an elastic body, where ${\bm r}$ denotes a spatial position.
In micropolar elasticity, each material point is allowed not only to translate but also to rotate independently. 
Therefore, we introduce a microrotation variable $\vb*{\phi}({\bm r})$ to describe this additional rotational degree of freedom associated with the material's microstructure.
For carbon nanotubes, the stretching mode can also be taken into account by considering an extension of micropolar theory, namely ``microstretch theory''.
In this theory, we further introduce a scalar microstretch $\Phi$, representing the isotropic volume change of a microelement.

We focus on longitudinal, torsional, and radial breathing modes in carbon nanotubes, for which the effective microstretch description simplifies.
We define the $x$ axis as the arc-length coordinate along the circumference, the $y$ axis along the tube axis, and the $z$ axis along the radial direction (see Fig.~\ref{coord_initial_twist}(a)). 
Since both the displacement and rotation axes are restricted to the nanotube axis, we need only consider $\phi_y$, $u_y$, and $\Phi$ as functions of the position variable $y$.
The variables $\phi_y$, $u_y$, and $\Phi$ correspond to the twisting mode, longitudinal acoustic mode, and radial breathing mode, respectively.
The kinetic energy density per unit length is given by
\begin{align}
    K_{\rm MSE}=\frac{1}{2}\rho j\dot{\phi}_y\dot{\phi}_y
    +\frac{1}{2}\rho\dot{u}_y\dot{u}_y
    +\frac{1}{2}\rho j_0\dot{\Phi}\dot{\Phi},
\end{align}
where $\rho$ is the mass density, $j$ is the microinertia, and $j_0$ is the microstretch inertia.
To align the dimensions of the variables, we define normalized variables, $\tilde{\phi}_y=\sqrt{j}\phi_y$ and $\tilde{\Phi}=\sqrt{j_0}\Phi$.
Then, the kinetic energy density becomes
\begin{align}
    K_{\rm MSE}=\frac{1}{2}\rho\dot{\tilde{\phi}}_y\dot{\tilde{\phi}}_y
    +\frac{1}{2}\rho\dot{u}_y\dot{u}_y
    +\frac{1}{2}\rho\dot{\tilde{\Phi}}\dot{\tilde{\Phi}}.
\end{align}

The potential energy density can be written as follows:
\begin{align}
    U_{\rm MSE}&=\frac{1}{2}A(\partial_y\tilde{\phi}_y)^2
    +\frac{1}{2}B(\partial_yu_y)^2
    +\frac{1}{2}C\tilde{\Phi}^2\nonumber\\
    &+D(\partial_y\tilde{\phi}_y)(\partial_yu_y)
    +E(\partial_yu_y)\tilde{\Phi}
    +F(\partial_y\tilde{\phi}_y)\tilde{\Phi},
    \label{UMSEeq}
\end{align}
where $A$, $B$, $C$, $D$, $E$, and $F$ are elastic constants.
Among the six terms on the right-hand side of Eq.~(\ref{UMSEeq}), $D(\partial_y\tilde{\phi}_y)(\partial_yu_y)$ and $F(\partial_y\tilde{\phi}_y)\tilde{\Phi}$ are referred to as chiral terms because they change sign under reflection in mirror planes perpendicular or parallel to the rod.
These chiral terms are allowed only when the one-dimensional rod possesses a chiral structure.
In the present case, the chiral terms appear only for twisted carbon nanotubes, as shown in Fig.~\ref{coord_initial_twist}(c).

The equation of motion can be derived from the Lagrangian density $\mathcal{L}_{\rm MSE}=K_{\rm MSE}-U_{\rm MSE}$.
Assuming $(\tilde{\phi}_y,u_y,\tilde{\Phi})=(\tilde{\phi}_y(k),u_y(k),\tilde{\Phi}(k))\exp[iky-i\omega(k)t]$, we obtain the following eigenvalue equation:
\begin{align}
    \omega(k)^2
    \mqty(\tilde{\phi}_y(k)\\u_y(k)\\\tilde{\Phi}(k))
    =\frac{1}{\rho}\mqty(Ak^2
    &Dk^2
    &-iFk\\
    Dk^2
    &Bk^2
    &-iEk\\
    iFk
    &iEk
    &C)
    \mqty(\tilde{\phi}_y(k)\\u_y(k)\\\tilde{\Phi}(k)).
    \label{eq:ABCDEF}
\end{align}
When $D=E=F=0$, the twisting and longitudinal modes, $\tilde{\phi}_y(k)$ and $u_y(k)$, are acoustic $(\omega(k) \propto k)$, whereas the stretch mode $\tilde{\Phi}$ is gapped $(\omega(k) \simeq {\rm const.})$.
The off-diagonal terms describe the mixing among these three modes.
The coupling of the stretch mode $\tilde{\Phi}(k)$ to the other two modes is proportional to $k$, whereas the coupling between $\tilde{\phi}_y$ and $u_y$ is proportional to $k^2$.
These features follow from the form of the couplings in the potential energy (\ref{UMSEeq}).

In the next section, we derive the chiral mixing terms proportional to $D$ and $F$ from a microscopic model of twisted carbon nanotubes.

\section{Microscopic formulation}
\label{sec:model}

We consider elastic waves in a twisted configuration of a carbon nanotube.
Since the modulation of the dispersion relation due to the initial twist does not occur within the framework of linear elasticity,
nonlinear terms must be taken into account~\cite{Pao1985}.
Therefore, we introduce nonlinear terms into the strain tensor in the elasticity theory of carbon nanotubes.

\subsection{Nonlinear strain tensor}
\label{sec:strain}

We describe carbon nanotube phonons within a continuum picture, treating a nanotube as a rolled-up graphene sheet~\cite{Saito1998}. Since graphene phonons behave as those of a two-dimensional isotropic elastic medium in the long-wavelength limit~\cite{Suzuura2002}, we model the nanotube as a cylindrical elastic sheet. 

In the undeformed configuration (see Fig.~\ref{coord_initial_twist}(b)), the position vector of a point on the cylindrical surface is written as
\begin{align}
    \vb*{p}
    =R\cos(x/R)\vb*{i}
    +R\sin(x/R)\vb*{j}
    +y\vb*{k},
    \label{p}
\end{align}
where $\vb*{i}$, $\vb*{j}$, and $\vb*{k}$ are the Cartesian unit vectors, and $R$ is the radius of the nanotube. After introducing the displacement field $\vb*{u}=(u_x,u_y,u_z)$, the deformed position is given by
\begin{align}
    \vb*{p}'
    &=(R+u_z)\cos[(x+u_x)/R]\vb*{i}\nonumber\\
    &+(R+u_z)\sin[(x+u_x)/R]\vb*{j}
    +(y+u_y)\vb*{k} .
    \label{p_deformed}
\end{align}
We then introduce the Green-Lagrange strain tensor $E_{\alpha\beta}$ on the cylindrical surface in terms of the change in the line element,
\begin{align}
    (\dd\vb*{p}')^2-(\dd\vb*{p})^2
    = 2E_{\alpha\beta}\dd{x}_\alpha\dd{x}_\beta.
\end{align}
Here and in what follows, repeated Greek indices are summed over $x$ and $y$. With $\dd\vb*{p}=(\partial_\alpha\vb*{p})\dd{x}_\alpha$ and $\dd\vb*{p}'=(\partial_\alpha\vb*{p}')\dd{x}_\alpha$, we obtain
\begin{align}
    E_{\alpha\beta}
    =\frac{1}{2}\qty[
    (\partial_\alpha\vb*{p}')
    \cdot(\partial_\beta\vb*{p}')
    -(\partial_\alpha\vb*{p})
    \cdot(\partial_\beta\vb*{p})].
\end{align}
Using Eqs.~\eqref{p} and \eqref{p_deformed} and retaining terms up to second order in the displacement field, we obtain
\begin{align}
    E_{xx}&=\partial_xu_x
    +u_z/R
    +2(\partial_xu_x)u_z/R\nonumber\\
    &+\frac{1}{2}[
    (u_z/R)^2
    +(\partial_xu_x)^2
    +(\partial_xu_y)^2
    +(\partial_xu_z)^2],\\
    E_{yy}&=\partial_yu_y
    +\frac{1}{2}[
    (\partial_yu_x)^2
    +(\partial_yu_y)^2
    +(\partial_yu_z)^2],\\
    E_{xy}
    &=\frac{1}{2}[
    \partial_xu_y
    +\partial_yu_x
    +2(\partial_yu_x)u_z/R
    +(\partial_xu_x)(\partial_yu_x)\nonumber\\
    &\hspace{2.5em}
    +(\partial_xu_y)(\partial_yu_y)
    +(\partial_xu_z)(\partial_yu_z)].
\end{align}
The first-order terms reproduce the conventional strain on a cylindrical surface, whereas the second-order terms describe the geometric nonlinearity associated with finite deformation. As shown in the next subsection, these second-order terms generate additional contributions to the linearized equation of motion around a twisted static configuration and are therefore essential for describing twist-induced mode coupling.

\subsection{Nonlinear equation of motion}

In this subsection, we derive the equation of motion for the displacement field $\vb*{u}$ on the cylindrical surface. The dynamics is described by the following Lagrangian density:
\begin{align}
    \mathcal{L} =\frac{1}{2}\rho\dot{u}_i\dot{u}_i  -U+b_iu_i,
\end{align}
where $\rho$ is the mass density and $b_i$ denotes an external body force density used below to sustain the statically twisted configuration of the nanotube.
Here and in what follows, repeated Latin indices are summed over $x$, $y$, and $z$.
The potential energy density is written in terms of the nonlinear strain tensor $E_{\alpha\beta}$ derived in Sec.~\ref{sec:strain} as
\begin{align}
    U=\frac{1}{2}C_{\alpha\beta\gamma\delta}
    E_{\alpha\beta}E_{\gamma\delta},
\end{align}
where $C_{\alpha\beta\gamma\delta}$ is the elastic tensor.
To isolate the terms relevant to the subsequent linearization around a twisted static configuration, we decompose the potential energy as $U=U_1+U_2$, where $U_1$ and $U_2$ are quadratic and cubic, respectively, in $\vb*{u}$, and neglect terms of fourth order and higher. We then obtain
\begin{widetext}
\begin{align}
    U_1
    &=\frac{1}{2}C_{\alpha\beta\gamma\delta}
    (\partial_\alpha u_\beta)
    (\partial_\gamma u_\delta)
    +C_{\alpha\beta xx}
    (\partial_\alpha u_\beta)
    (u_z/R)+\frac{1}{2}C_{xxxx}(u_z/R)^2, \\
    U_2
    &=\frac{1}{2}C_{\alpha\beta\gamma\delta}
    (\partial_\alpha u_\beta)
    (\partial_\gamma u_i)
    (\partial_\delta u_i)
    +2C_{\alpha\beta xx}
    (\partial_\alpha u_\beta)
    (\partial_xu_x)u_z/R
    +2C_{\alpha\beta xy}
    (\partial_\alpha u_\beta)
    (\partial_yu_x)u_z/R \nonumber\\
    &+\frac{1}{2}C_{\alpha\beta xx}
    (\partial_\alpha u_i)
    (\partial_\beta u_i)
    u_z/R
    +\frac{1}{2}C_{\alpha\beta xx}
    (\partial_\alpha u_\beta)
    (u_z/R)^2
    +2C_{xxxx}
    (\partial_xu_x)(u_z/R)^2\nonumber\\
    &+2C_{xxxy}
    (\partial_yu_x)(u_z/R)^2
    +\frac{1}{2}C_{xxxx}
    (u_z/R)^3.
\end{align}
\end{widetext}

The equation of motion then follows from the Euler-Lagrange equation
\begin{align}
    \partial_t\pdv{\mathcal{L}}{\dot{u}_i}
    +\partial_\alpha\pdv{\mathcal{L}}{(\partial_\alpha u_i)}
    -\pdv{\mathcal{L}}{u_i}=0,
\end{align}
which gives
\begin{align}
    \rho\ddot{u}_i
    &=f_i[\vb*{u}]
    +b_i, \\
    f_i[\vb*{u}]
    &=\partial_\alpha\pdv{U}{(\partial_\alpha u_i)}
    -\pdv{U}{u_i}.
\end{align}
The elastic force can be decomposed as $f_i[\vb*{u}]=f^{(1)}_i[\vb*{u}]+f^{(2)}_i[\vb*{u}]$, where
\begin{align}
    f^{(d)}_i[\vb*{u}]
    =\partial_\alpha\pdv{U_d}{(\partial_\alpha u_i)}
    -\pdv{U_d}{u_i},
    \qquad (d=1,2).
\end{align}
The linear contributions
$f^{(1)}_\alpha[\vb*{u}]~
(\alpha=x,y)$ and
$f^{(1)}_z[\vb*{u}]$ are given by
\begin{align}
    f^{(1)}_\alpha[\vb*{u}]
    =C_{\alpha\beta\gamma\delta}
    \partial_\beta\partial_\gamma u_\delta
    +C_{\alpha\beta xx}R^{-1}
    \partial_\beta u_z,
\end{align}
\begin{align}
    f^{(1)}_z[\vb*{u}]
    =-C_{\alpha\beta xx}R^{-1}
    \partial_\alpha u_\beta
    -C_{xxxx}R^{-2}u_z.
\end{align}
The quadratic contributions
$f^{(2)}_\alpha[\vb*{u}]~
(\alpha=x,y)$ and
$f^{(2)}_z[\vb*{u}]$ are given by
\begin{widetext}
\begin{align}
    f^{(2)}_\alpha[\vb*{u}]
    &=\frac{1}{2}C_{\alpha\beta\gamma\delta}
    \partial_\beta((\partial_\gamma u_i)
    (\partial_\delta u_i))
    +C_{\mu\nu\gamma\beta}
    \partial_\beta((\partial_\mu u_\nu)
    (\partial_\gamma u_\alpha))
    +2C_{\alpha\beta xx}
    \partial_\beta((\partial_xu_x)u_z/R)
    +2C_{\mu\nu xx}
    \delta_{\alpha x}
    \partial_x
    ((\partial_\mu u_\nu)
    u_z/R)&\nonumber\\
    &+2C_{\alpha\beta xy}
    \partial_\beta((\partial_yu_x)u_z/R)
    +2C_{\mu\nu xy}
    \delta_{\alpha x}
    \partial_y
    ((\partial_\mu u_\nu)
    u_z/R)
    +C_{\mu\beta xx}
    \partial_\beta
    ((\partial_\mu u_\alpha)
    u_z/R)
    +2C_{\alpha\beta xx}R^{-2}
    u_z(\partial_\beta u_z)  \nonumber \\
    &+4C_{xxxx}
    \delta_{\alpha x}
    u_z(\partial_x u_z)
    +4C_{xxxy}
    \delta_{\alpha x}
    u_z(\partial_y u_z), \\
    f^{(2)}_z[\vb*{u}]
    &=C_{\mu\nu\gamma\beta}
    \partial_\beta
    ((\partial_\mu u_\nu)
    (\partial_\gamma u_z))
    +C_{\alpha\beta xx}
    \partial_\beta
    ((\partial_\alpha u_z)
    u_z/R)
    -2C_{\alpha\beta xx}R^{-1}
    (\partial_\alpha u_\beta)
    (\partial_xu_x)
    -2C_{\alpha\beta xy}R^{-1}
    (\partial_\alpha u_\beta)
    (\partial_yu_x)
    \nonumber\\
    &-\frac{1}{2}C_{\alpha\beta xx}R^{-1}
    (\partial_\alpha u_i)
    (\partial_\beta u_i)
    -C_{\alpha\beta xx}R^{-2}
    (\partial_\alpha u_\beta)
    u_z -4C_{xxxx}R^{-2}
    (\partial_xu_x)u_z
    -4C_{xxxy}R^{-2}
    (\partial_yu_x)u_z\nonumber\\
    &-\frac{3}{2}C_{xxxx}R^{-3}
    u_z^2.
\end{align}
\end{widetext}
For the undeformed nanotube, the equilibrium configuration is $\vb*{u}=0$, and the linearized equation of motion contains only $f_i^{(1)}[\vb*{u}]$. By contrast, if the equilibrium state is a finite static deformation $\vb*{u}^{\rm i}$, the expansion of $f_i^{(2)}[\vb*{u}^{\rm i}+\vb*{u}]$ contains terms linear in the fluctuation $\vb*{u}$, which modify the linearized equation of motion. We analyze this mechanism explicitly in the next subsection for a uniformly twisted configuration.

\subsection{Linearized equation of motion around a twisted configuration}

We consider a uniformly twisted static configuration defined by
$u^{\rm i}_x=\theta y$, $u^{\rm i}_y=u^{\rm i}_z=0$,
where $\theta$ is a constant twist rate.
The equilibrium equation, with an external body force $b_i$ applied to sustain this deformation, is given by
\begin{align}
    0
    =f^{(1)}_i[\vb*{u}^{\rm i}]
    +f^{(2)}_i[\vb*{u}^{\rm i}]
    +b_i.
    \label{eq:twisted_equilibrium}
\end{align}
We then introduce fluctuations around this static configuration by writing the displacement field as
$\vb*{u}^{\rm i}(x,y)+\vb*{u}(x,y,t)$.
The corresponding equation of motion is
\begin{align}
    \rho\ddot{u}_i
    =f^{(1)}_i[\vb*{u}^{\rm i}+\vb*{u}]
    +f^{(2)}_i[\vb*{u}^{\rm i}+\vb*{u}]
    +b_i.
    \label{eq:equilibriumEOM}
\end{align}
Subtracting the equilibrium equation \eqref{eq:twisted_equilibrium} from \eqref{eq:equilibriumEOM} eliminates $b_i$ and yields the equation of motion for the fluctuation $\vb*{u}$ around $\vb*{u}^{\rm i}$:
\begin{align}
    \rho\ddot{u}_i
    =f^{(1)}_i[\vb*{u}]
    +f^{(2)}_i[\vb*{u}^{\rm i}+\vb*{u}]
    -f^{(2)}_i[\vb*{u}^{\rm i}].
    \label{eq:EOM_around_twisted_configuration}
\end{align}
Here, we have used the linearity of $f^{(1)}_i[\vb*{u}]$ with respect to $\vb*{u}$, which implies
$f^{(1)}_i[\vb*{u}^{\rm i}+\vb*{u}]
-f^{(1)}_i[\vb*{u}^{\rm i}]
=f^{(1)}_i[\vb*{u}]$.
In the following, we retain only the terms linear in the fluctuation $\vb*{u}$ in
$f^{(2)}_i[\vb*{u}^{\rm i}+\vb*{u}]
-f^{(2)}_i[\vb*{u}^{\rm i}]$.
To obtain an explicit form of the equation of motion, we adopt the isotropic elastic approximation for the graphene sheet in the long-wavelength limit. The elastic tensor is then expressed in terms of the Lam\'e constants $\lambda$ and $\mu$ as
\begin{align}
    C_{\alpha\beta\gamma\delta}
    =\lambda\delta_{\alpha\beta}\delta_{\gamma\delta}
    +\mu(\delta_{\alpha\gamma}\delta_{\beta\delta}
    +\delta_{\alpha\delta}\delta_{\beta\gamma}).
\end{align}
Finally, we obtain
\begin{widetext}
\begin{align}
    f^{(2)}_x[\vb*{u}^{\rm i}+\vb*{u}]
    -f^{(2)}_x[\vb*{u}^{\rm i}]
    & =2(\lambda+2\mu)\theta
    \partial_x\partial_yu_x
    +\mu\theta
    \partial_x\partial_xu_y
    +(\lambda+2\mu)\theta
    \partial_y\partial_yu_y
    +(\lambda+4\mu)\theta R^{-1}
    \partial_yu_z, \\
    f^{(2)}_y[\vb*{u}^{\rm i}+\vb*{u}]
    -f^{(2)}_y[\vb*{u}^{\rm i}]
    &=\mu\theta
    \partial_x\partial_xu_x
    +(\lambda+2\mu)\theta
    \partial_y\partial_yu_x
    +2\mu\theta
    \partial_x\partial_yu_y
    +2\mu\theta R^{-1}
    \partial_xu_z, \\
    f^{(2)}_z[\vb*{u}^{\rm i}+\vb*{u}]
    -f^{(2)}_z[\vb*{u}^{\rm i}]
    &=-(\lambda+4\mu)\theta R^{-1}
    \partial_yu_x
    -2\mu\theta R^{-1}
    \partial_xu_y
    +2\mu\theta
    \partial_x\partial_yu_z.
\end{align}
\end{widetext}

\subsection{Dynamical matrix}

Substituting the plane-wave solution propagating along the tube axis,
\begin{align}
    \vb*{u}(x,y,t)=\vb*{u}(k)e^{iky-i\omega(k)t},
\end{align}
into Eq.~\eqref{eq:EOM_around_twisted_configuration}, we obtain the eigenvalue equation
\begin{align}
    \omega(k)^2\mqty(u_x(k)\\u_y(k)\\u_z(k))
    =\mathcal{D}(k)\mqty(u_x(k)\\u_y(k)\\u_z(k)),
    \label{eigenequation}
\end{align}
where the dynamical matrix is given by
\begin{align}
    & \mathcal{D}(k) \nonumber \\
    &=\frac{1}{\rho}\mqty(\mu k^2&(\lambda+2\mu)\theta k^2
    &-i(\lambda+4\mu)\theta R^{-1}k\\
    (\lambda+2\mu)\theta k^2&(\lambda+2\mu)k^2&-i\lambda R^{-1}k\\
    i(\lambda+4\mu)\theta R^{-1}k&i\lambda R^{-1}k&(\lambda+2\mu)R^{-2}).\label{Dmatrix}
\end{align}
Here, $u_x$, $u_y$, and $u_z$ represent the twisting, longitudinal acoustic, and radial breathing components, respectively.

Eqs.~\eqref{eigenequation} and \eqref{Dmatrix} can be identified with the effective one-dimensional microstretch theory introduced in Sec.~\ref{sec:microstretch}.
In that correspondence, the microstretch variables $\tilde{\phi}_y$, $u_y$, and $\tilde{\Phi}$ are identified with $u_x$, $u_y$, and $u_z$, respectively.
The coefficients $A$, $B$, $C$, $D$, $E$, and $F$ in Eq.~\eqref{eq:ABCDEF} are determined as
\begin{subequations}
\begin{align}
    A&=\mu, \\
    B&=\lambda + 2\mu, \\
    C&=(\lambda + 2\mu) R^{-2}, \\
    D&=(\lambda+2\mu)\theta,\\
    E&=\lambda R^{-1}, \\
    F&=(\lambda+4\mu)\theta R^{-1}.
\end{align}
\end{subequations}
The elastic constants $D$ and $F$ are proportional to the twist rate $\theta$.
Thus, the twisted equilibrium configuration generates the chiral terms and induces mixing between the twisting mode and the longitudinal and breathing modes. This establishes a direct microscopic connection between structural chirality and the effective microstretch description of phonons in carbon nanotubes, and constitutes the central result of the present work.

\subsection{Dispersion relation}

\begin{figure}
    \centering
    \includegraphics[width=8cm]{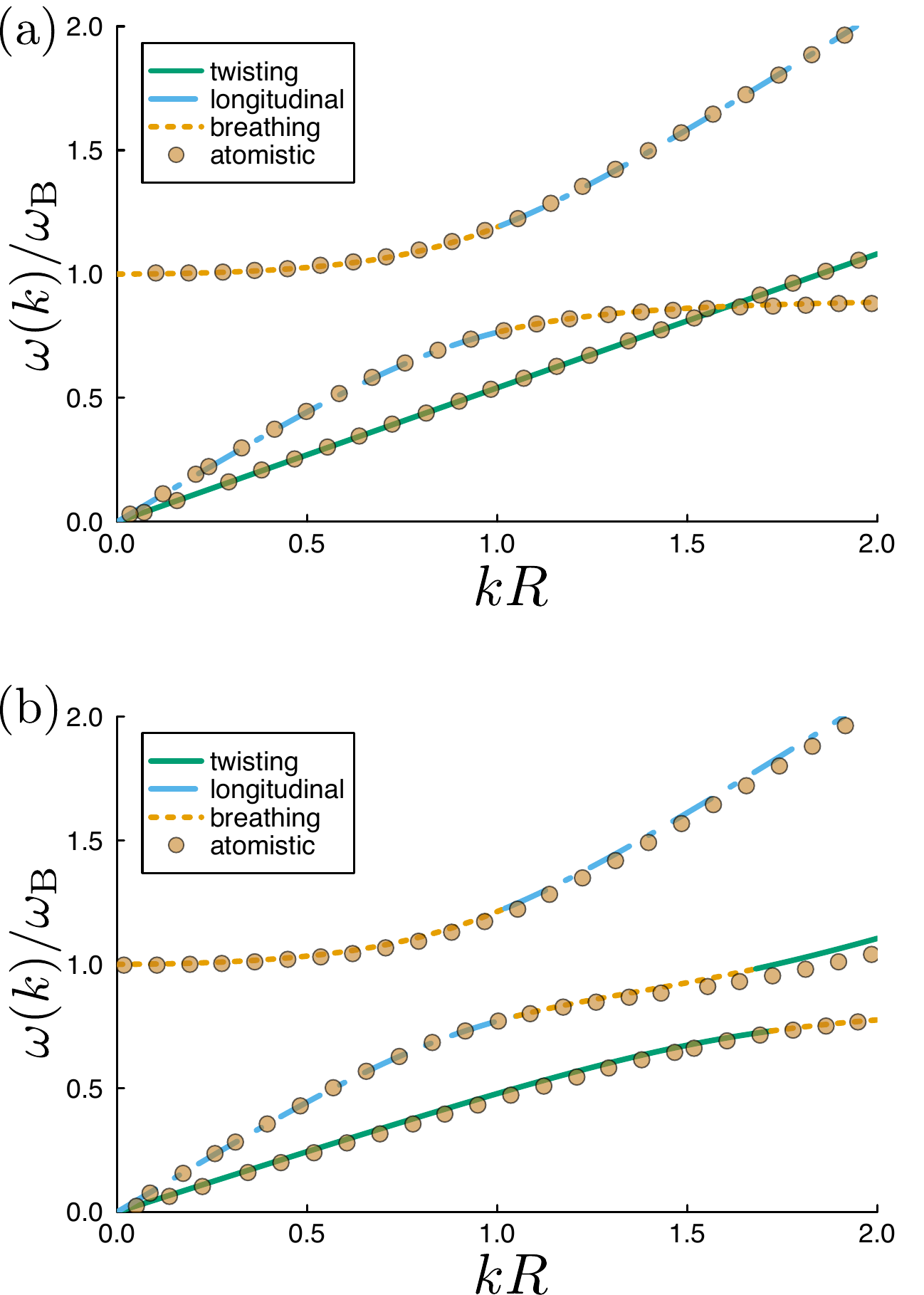}
    \caption{Dispersion relations of the twisting, longitudinal, and breathing modes.
    The eigenstates are hybridized in general and are color-coded according to the component with the largest weight.
    The frequency is normalized by the breathing-mode frequency $\omega_{\rm B}=R^{-1}\sqrt{(\lambda+2\mu)/\rho}$.
    The Lam\'e-constant ratio is set to $\lambda/\mu=1.42$.
    The circles indicate atomistic data from Ref.~\cite{GuptaKumar2019}.
    (a) Untwisted nanotube ($\theta=0$).
    (b) Twisted nanotube ($\theta=0.139$).}
    \label{dispersion}
\end{figure}

Finally, we briefly discuss the features of the phonon dispersion.
The solid and dashed lines in Fig.~\ref{dispersion} show the phonon dispersion relations obtained from the dynamical matrix given in Eq.~\eqref{Dmatrix}.
The parameters $\lambda$, $\mu$, and $R$ are chosen to fit the molecular-dynamics results~\cite{GuptaKumar2019}, which are denoted by the filled circles.
For the untwisted nanotube ($\theta=0$), the twisting mode is decoupled from the longitudinal and breathing modes, as shown in Fig.~\ref{dispersion}(a).
In contrast, for a twisted nanotube ($\theta=0.139$), the twist-induced chiral coupling hybridizes the twisting and breathing modes and opens an anticrossing gap, as shown in Fig.~\ref{dispersion}(b).
Notably, our approach provides a satisfactory fit to the molecular-dynamics data with fewer parameters than those required in the phenomenological microstretch theory.

Since our theory includes only three independent parameters, $\lambda$, $\mu$, and $R$, the parameters $A$, $B$, $C$, $D$, $E$, and $F$ are not independent. 
To see this, we consider the following dimensionless parameters:
\begin{subequations}
\begin{align}
    A/(CR^2)&=\mu/(\lambda+2\mu),\\
    B/(CR^2)&=1,\\
    D/(CR^2)&=\theta,\\
    E/(CR)&=\lambda/(\lambda+2\mu),\\
    F/(CR)&=(\lambda+4\mu)\theta/(\lambda+2\mu).
\end{align}
\end{subequations}
In our calculation, $B/(CR^2) = 1$ and $D/(CR^2)=\theta$ always hold, whereas these relations need not hold in the microstretch elastic theory based on fitting to the molecular-dynamics data~\cite{GuptaKumar2019}.
We compare these dimensionless parameters between our calculation ($\lambda/\mu=1.42$) and Ref.~\cite{GuptaKumar2019} for the untwisted and twisted carbon nanotubes in Tables~\ref{tb:params1} and \ref{tb:params2}, respectively.
The prediction $B/(CR^2) = 1$ is well reproduced by the molecular-dynamics data.
For the other parameters, while we find good agreement for untwisted nanotubes, deviations are observed clearly for twisted nanotubes.
These deviations are expected to arise from curvature effects or anisotropy in the elastic tensor.
Overall, the agreement between our calculation and the molecular-dynamics results is satisfactory considering the simplicity of our formulation.

\begin{table}[tb]
    \centering
    \begin{tabular}{llllll}
         &$A/CR^2$& $B/CR^2$ & $D/CR^2$ & $E/CR$ & $F/CR$ \\\hline
        Ref.~\cite{GuptaKumar2019} & 0.292 & 0.963 & 0.000 & 0.448 & 0.000 \\
        Our calculation & 0.292 & 1.000 & 0.000 & 0.415 & 0.000 
    \end{tabular}
    \caption{Parameters for the untwisted nanotubes.}
    \label{tb:params1}
\end{table}

\begin{table}[tb]
    \centering
    \begin{tabular}{llllll}
         &$A/CR^2$& $B/CR^2$ & $D/CR^2$ & $E/CR$ & $F/CR$ \\\hline
        Ref.~\cite{GuptaKumar2019} & 0.220 & 0.980 & -0.054 & 0.459 & 0.117 \\
        Our calculation & 0.292 & 1.000 & 0.139 & 0.415 & 0.221
    \end{tabular}
    \caption{Parameters for the twisted nanotubes.}
    \label{tb:params2}
\end{table}

\section{Summary}
\label{sec:summary}

We have provided a microscopic derivation of an effective one-dimensional microstretch description for phonons in twisted carbon nanotubes. In these systems, the relevant low-energy modes include not only a torsional mode, naturally associated with microrotation, but also the radial-breathing mode, which requires an additional scalar stretch degree of freedom. Accordingly, the effective phonon dynamics is captured by a microstretch theory.

Starting from nonlinear elasticity on a cylindrical surface, we derived the equation of motion for the displacement field and linearized it around a uniformly twisted static configuration. We showed that the chiral coupling terms of the effective theory arise from this twisted equilibrium configuration through the linearization of the nonlinear elastic force. The resulting dynamical matrix for the twisting, longitudinal, and radial-breathing modes coincides with that of a one-dimensional microstretch theory. Within an isotropic elastic approximation, we further obtained analytic expressions for the effective coefficients in terms of the Lam\'e constants, the nanotube radius, and the twist rate.
The twist-induced chiral couplings hybridize the twisting, longitudinal, and breathing modes and open an anticrossing in the phonon dispersion.

Our study illustrates how generalized continuum theories can be connected to microscopic phonon dynamics in chiral elastic systems.
Extending the present approach to derive generalized continuum theories microscopically for broader classes of chiral materials remains an important problem for future work.

\begin{acknowledgments}
We would like to thank T. Yamamoto, R. Akimoto, H. Matsuura, W. Izumida, A. Yamakage, T. Funato, H. Funaki, R. Sano, Y. Sekino, and H. Tajima for their valuable discussions.  
We appreciate the insightful suggestion by J. Kishine regarding the fundamental link between spin-mechatronics and Eringen's micropolar theory. 
This work was supported by the National Natural Science Foundation of China (NSFC) under Grant No. 12374126, by the Priority Program of Chinese Academy of Sciences under Grant No. XDB28000000, and by JSPS KAKENHI for Grants (Nos. JP23H01839, JP24H00322, and JP24K06951) from MEXT, Japan. 
N.N. was supported by Forefront Physics and Mathematics Program to Drive Transformation (FoPM), a World-leading Innovative Graduate Study (WINGS) Program, the University of Tokyo.
\end{acknowledgments}

\bibliography{ref}

\end{document}